\documentclass[review, pdftex]{elsarticle}
\usepackage{setspace}
\usepackage{amsmath}
\usepackage{amssymb}
\usepackage{graphicx}
\usepackage{color}

\journal{Combustion Symposium}

\begin{document}
\begin{frontmatter}
\title{A mathematical model for three dimensional detonation as pure
gas-dynamic discontinuity}

\author{Jorge Y\'a\~nez Escanciano\corref{cor1}}
\cortext[cor1]{Corresponding Author}
\ead{jorge.yanez@kit.edu}

\author{Andreas G. Class\corref{cor2}}

\address{Institute for Energy and Nuclear Technique, Karlsruhe Institute of Technology,\\[0.4em] Hermann-von-Helmholtz-Platz 1, 76344 Eggenstein-Leopoldshafen, Germany}

\begin{abstract}
A model for three dimensional detonation is proposed based on the approximation that the detonation thickness is small compared to the characteristic scales of the fluid motion. In this framework detonations are treated as a modified hydrodynamic discontinuity. The altered Rankine-Hugoniot jump conditions take into account the internal structure of the detonation including the chemical reaction. The position of the discontinuity surface and the corresponding jump conditions are derived from first principles. The final modified conditions are dependent on curvature, flame thickness and stretching and allow for simple physical interpretation. 
\end{abstract}


\end{frontmatter}
{\bf Additional information for reviewers:}
\begin{itemize}
\item Colloquium intended: Detonation explosions and supersonic combustion.

\item Total legth of the paper and method of determination: 5706 words excluding, as stated in the instructions for authors, title block, abstract and separate list of figures and captions. Methodology number 2 special for Latex users. In the preamble options 5p and twocolumn were selected.

\item Equivalent Legths: Main Text: 4493, Nomenclature: 277, References: 283, Figure 1: 190, Figure 2: 147, Figure 3: 157, Figure 4: 159

\end{itemize}

\newpage

\section*{Nomenclature}
\begin{footnotesize} %
\begin{tabular}{lll}
Latin & & Greek\\
\hline
\begin{tabular}{p{0.2cm}p{2.8cm}}
$a$ & Auxiliary variable\\
$b$ & Auxiliary variable\\
$c$ & Sound speed\\
$D$ & Detonation speed\\
$\overline{\overline{E}}$ & Unit Tensor\\
$E$ & Energy, reac. model\\
$E_{a}$ & Activation energy\\
$e$ & Energy, hydr. model\\
$g$ & Metric tensor\\
$I$ & Integral\\
$J$ & Flux\\
$H$ & Average curvature\\
$k$ & Pre-exponential factor\\
$L$ & Progress variable, reac. model\\
$l$ & Parallel vector fields\\
$l_{f}$ & Scale flow motion\\
\end{tabular} & %
\begin{tabular}{p{0.2cm}p{2.8cm}}
$l_{c}$ & Half reac. zone\\
$M$ & Mach number\\
$m$ & Momentum\\
$p$ & Pressure\\
$Q$ & Source term\\
$R$ & Density, reac. model\\
$R_{g}$ & Constant of gases\\
$T$ & Temperature \\
$t$ & Time\\
$u$ & Speed, moving coord.\\
$V$ & Speed, reac. model\\
$v$ & Speed, hydr. model\\
$W$ & Consumption rate\\
$x$ & Position\\
$X$ & Position, stretched coord.\\
$Y$ & Progress variable\\
$Z$ & Ratio $\tilde{l_f}/\tilde{l_{c}}$\\
\end{tabular} & %

\begin{tabular}{p{0.2cm}p{2.8cm}}
$\alpha$ & Surface tension coef.\\
$\beta$ & Dimensionless enth. formation\\
$\gamma$ & Ratio specific heats\\
$\zeta$ & Progress variable\\
$\theta$ & Auxiliary variable\\
$\Theta$ & Auxiliary variable\\
$\theta_{a}$ & Dimensionless act. energy\\
$\lambda$ & Progress variable\\
$\rho$ & Density\\
$\Upsilon$ & Specific volume\\
$\phi$ & Hydrodynamic model var.\\
$\Phi$ & Reactive model var.\\
$\chi$ & Stretching\\
\end{tabular}\\

\end{tabular}\end{footnotesize}

\section{Introduction}
\label{sec:introduction} 

Accidents involving the release and subsequent combustion of burnable gases can be classified, utilizing the flame propagation regime, ranging from slow deflagrations to detonations. The latter appears to be particularly important since detonations are usually considered as a \emph{worst case}, often resulting in accident scenarios with devastating consequences.

During the last decades numerous modeling studies (e.g., \cite{Lee08}, \cite{She09}, \cite{Ora09}, \cite{Wil96}) showed that the detonation process in combustible gaseous mixtures can be successfully reproduced if the internal structure of the detonation wave is resolved. Good practice implies resolving the 3D structure of detonation cells, but ignoring details of the internal structure of the shock or of the chemistry \cite{Ora09}, \cite{Wil96}.

The thickness of detonations is typically small compared to the characteristic scales of the fluid flow. Jin and Stewart \cite{Jin96} and Klein et al. \cite{Kle95} utilized the limit of asymptotically thin detonations to derive the asymptotic theory for weakly curved detonations. Thus, a simplified model for detonations can be built where the fuel consumption zone shrinks to an infinitely thin surface of discontinuity separating the cold mixture from the hot products. For deflagrations, e.g. laminar premixed combustions, models considering flames as a gas-dynamic discontinuity exist since the pioneering works of Darrieus and Landau who derived the jump conditions across the flame. More recently, Matalon and Matkowsky \cite{Mat82} considered arbitrary flame shapes for nearly equi-diffusional flames with thermal expansion in general flow fields. The leading terms of the jump conditions were those of the Darrieus-Landau model and perturbative corrections were obtained in the next order of approximation. Their formulation is based on high activation energy asymptotics. Klimenko and Class \cite{Kli00} employed tensor calculus and orthonormal coordinates to simplify the derivation of the flame speed relation of Matalon and Matkowsky. Their approach explicitly exploits the distinctiveness of length scales instead of high activation energy, i.e. large Peclet number $Pe$, defined as the ratio of flow length scale to flame thickness. In the subsequent series of papers, their methodology was applied to multi-step chemistry \cite{Kli02} and wider reaction zones \cite{Kli02b}. Finally, Class et al. \cite{Cla03} reconsidered the Rankine-Hugoniot jump conditions for the flow field and showed that perturbative correction of the jump conditions depend on perturbative corrections of the location where these jump conditions are evaluated. Moreover, they shown that there exists a unique location within the flame structure, where no extra inertia terms arise due to the continuity surface. This corresponds to a vanishing perturbative correction of surface mass. 

In the present work, the authors apply the Class et al. \cite{Cla03} methodology to detonation. Modified Rankine-Hugoniot jump conditions are derived, implicitly including the full effect of the chemistry. The methodology developed by Class et al. \cite{Cla03} has the essential advantage that perturbative corrections of the jump conditions are expressed as functions of the leading order solutions, so that the explicit evaluation of perturbative corrections of the solution becomes outdated. For detonation, this implies that the results of the Zeldovich-von Newmann-D\"oring theory can be utilized as a leading order \emph{planar} model to close the system \cite{Lee08}, \cite{Fic79} and obtain the final modified jump conditions which include perturbative correction terms.
\section{Analysis}
\label{sec:ana}
\subsection{Governing equations}
\label{subsec:gobern} The reactive Euler equations are considered,
\begin{align}
\partial_{t}\tilde{\phi}+\nabla\cdot J(\tilde{\phi}) &
=\tilde{Q}(\tilde{\phi}),
\end{align}
where for the continuity, momentum, energy and species
\begin{equation}
\tilde{\phi}=\begin{pmatrix}\tilde{\rho}\\
\tilde{\rho}\tilde{v}\\
\tilde{\rho}\tilde{e}\\
\tilde{\rho}\tilde{Y}
\end{pmatrix},\; J\begin{pmatrix}\tilde{\rho}\\
\tilde{\rho}\tilde{v}\\
\tilde{\rho}\tilde{e}\\
\tilde{\rho}\tilde{Y}
\end{pmatrix}=\begin{pmatrix}\tilde{\rho}\tilde{v}\\
\tilde{\rho}\tilde{v}\otimes\tilde{v}+\tilde{p}\overline{\overline{E}}\\
\tilde{\rho}\tilde{e}\tilde{v}+\tilde{p}\overline{\overline{E}}\cdot\tilde{v}\\
\tilde{\rho}\tilde{Y}\tilde{v}
\end{pmatrix},\; Q\begin{pmatrix}\tilde{\rho}\\
\tilde{\rho}\tilde{v}\\
\tilde{\rho}\tilde{e}\\
\tilde{\rho}\tilde{Y}
\end{pmatrix}=\begin{pmatrix}0\\
0\\
0\\
\tilde{\rho}\tilde{W}
\end{pmatrix}.
\end{equation}
The system can be made dimensionless with the use of the reference variables $\tilde{\rho}_{s}$, $\tilde{c}_{s}$, $\tilde{l}_{f}$ where the sub index $s$ represents the conditions at the Von Neumann peak and $\tilde{l}_{f}$ is the characteristic scale of the flow motion. The dependent variables can be written as $\tilde{\rho}=\rho\tilde{\rho}_{s},$ $\tilde{v}=v\tilde{c}_{s},$ $\tilde{p}=p\cdot\tilde{\rho}_{s}\tilde{c}_{s}^{2}/\gamma,$ $\tilde{e}=e\tilde{c}_{s}^{2}$, $\tilde{Y}=\lambda\tilde{Y_{s}}$ while the independent variables become $\tilde{t}=t\cdot\tilde{l}_{f}/\tilde{c}_{s},$ $\tilde{x}=x\tilde{l}_{f}.$ The Arrhenius chemical consumption rate can be written using the variables $\theta_{a}=\tilde{E_{a}}/(\tilde{R_{g}}\tilde{T_{s}})$ and $k=(\tilde{k}\tilde{l_{c}})/\tilde{c}_{s})$ as 
\begin{align}
W & =k(1-Y)\exp{\left(-\theta_{a}\rho/p\right)}.
\end{align}

We assume a large ratio $Z$  of the hydrodynamic typical length to the half detonation thickness $Z=\tilde{l}_{f}/\tilde{l_{c}}$. A thin detonation structure corresponds to an intense source term $Q$ which is re-scaled accordingly $Q=Z/Z\cdot Q=ZQ'$ yielding 
\begin{align}
\partial_{t}\phi+\nabla\cdot J(\phi) & =ZQ'(\phi).
\end{align}

It is convenient to transform the equations to a moving generalized curvilinear coordinate system. These coordinates were utilized in the references \cite{Kli00}, \cite{Kli02}, \cite{Kli02b}, \cite{Cla03}. The coordinate system is attached to the discontinuity surface with its normal direction pointing towards the products. Its tangential direction moves with the local tangential flow. In this system, the flame is at rest with no flow along the flame surface. The tensor calculus utilized in our analysis is mainly contained in reference \cite{Ari89}. The coordinates \cite{Cla03} are orthonormal with $x^{1}$ direction perpendicular to the surface of the flame, so that  $x^{1}$ coordinate lines are normal to the surfaces $x^{1}=const$. The contra-variant metric tensor is given by $g^{ij}$. Due to the orthogonality $g^{1\alpha}=0$, and due to normalization $g^{11}=1$. In the curvilinear coordinates, the system of differential equations can be written as
\begin{align}
\partial_{t}\left(\sqrt{g}\phi\right)+\partial_{x^{j}}\left(\sqrt{g}J^{j}(\phi)\right)
& =\sqrt{g}ZQ'(\phi)\label{eq:euler_ini}
\end{align}
with the flux vectors as $J^{j}(\phi)=(v^{j}-u^{j})\phi$ with $u^{j}$ representing the speed of the moving coordinates relative to fixed Eulerian coordinates. The fluxes become according to \cite{Cla03}, 
\begin{align}
J^{j}\begin{pmatrix}\rho\\
\rho v^{i}l_{i}\\
\rho e\\
\rho\lambda
\end{pmatrix}=\begin{pmatrix}m^{j}\\
\left(m^{j}v^{i}+p/\gamma\cdot g^{ij}\right)l_{i}\\
m^{j}e+p/\gamma\cdot g^{ij}\cdot m^{i}/\rho\\
m^{j}\lambda
\end{pmatrix},
\end{align}
with mass flux $m^{j}=(v^{j}-u^{j})\rho$. Decomposing the Eq. \eqref{eq:euler_ini} in the normal and tangential directions and introducing the stretched normal spatial variable, $X=Zx^{1}$, Eq. \eqref{eq:euler_ini} yields,
\begin{align}
Z^{-1}\left[\partial_{t}\left(\sqrt{g}\phi\right)+\partial_{x^{\alpha}}\left(\sqrt{g}J^{\alpha}(\phi)\right)\right]+\partial_{X}\left(\sqrt{g}J^{1}(\phi)\right)
& =\sqrt{g}Q'(\phi).\label{eq:all_junt}
\end{align}
\subsection{Asymptotic derivation of the fluid equations and jump conditions}
\label{subsec:asimpt}
Since $Z$, the ratio between the length of the fluid flow and the consumption zone thickness, is assumed to be asymptotically large, the  variables are expressed in terms of an asymptotic series expansion in powers of $1/Z$ , i.e. $\phi\approx\sum_{n=0}Z^{-n}\phi_{(n)}\approx\phi_{(0)}+Z^{-1}\phi_{(1)}+O(Z^{-1})$. The volume element is Taylor expanded around the discontinuity $\sqrt{g}=\sqrt{g}_{(0)}+\sqrt{g}_{(1)}Z^{-1}X+o(X^{2})$. Additionally, we make use of the equalities $\sqrt{g}_{(1)}=-2H$ and $\chi=\partial_{t}(\sqrt{g}_{(0)})/\sqrt{g}_{(0)}$ with mean curvature $H$ and stretch $\chi$. Equation \eqref{eq:all_junt} becomes
\begin{align}
& Z^{-1}\left[\left(\partial_{t}+\chi\right)(\phi_{(0)}+Z^{-1}\phi_{(1)})+(g_{(0)})^{-1/2}\partial_{x^{\alpha}}\left(\sqrt{g}_{(0)}(J_{(0)}^{\alpha}(\phi)+Z^{-1}J_{(1)}^{\alpha})\right)\right]+\nonumber\\
&+\partial_{X}\left[\left(1-Z^{-1}2HX\right)(J_{(0)}^{1}(\phi)+Z^{-1}J_{(1)}^{1})\right]=\left(1-Z^{-1}2HX\right)(Q'_{(0)}(\phi)+Z^{-1}Q'_{(1)}).
\end{align}
Collecting coefficients of like powers in $Z^{-n}$, the zeroth, $Z^{0}$, and first order $Z^{-1}$ terms are 
\begin{align}
& \partial_{X}\left(J_{(0)}^{1}(\phi)\right)=Q'_{(0)}(\phi),\label{eq:J0}\\
&\left(\partial_{t}+\chi\right)\phi_{(0)}+(g_{(0)})^{-1/2}\partial_{x^{\alpha}}\left(\sqrt{g}_{(0)}J_{(0)}^{\alpha}(\phi)\right)+\nonumber\\
&+\partial_{X}\left(J_{(1)}^{1}-2HXJ_{(0)}^{1}(\phi)\right)=Q'_{(1)}(\phi)-2HXQ'_{(0)}(\phi).\label{eq:J1}
\end{align}
The leading order normal fluxes are,
\begin{align}
J_{(0)}^{1}\begin{pmatrix}\rho\\
\rho v^{i}l_{i}\\
\rho e\\
\rho\lambda
\end{pmatrix}=\begin{pmatrix}m_{(0)}^{1}\\
m_{(0)}^{1}\left(v_{(0)}^{1}l_{1_{(0)}}+v_{(0)}^{\alpha}l_{\alpha_{(0)}}\right)+p_{(0)}/\gamma\cdot
l_{1_{(0)}}\\
m_{(0)}^{1}e_{(0)}+p_{(0)}/\gamma\cdot g_{(0)}^{i1}\cdot
m_{(0)}^{i}/\rho_{(0)}\\
m_{(0)}^{1}\lambda_{(0)}
\end{pmatrix}.\label{eq:flux0}
\end{align}
The fluxes in the momentum equation can be decomposed into the terms emanating from the tangential and normal components. Using a parallel vector field $l_{i}$ normal to the flame at the point of consideration, 
\begin{align}
J_{(0)}^{1}(\rho v^{i}l_{\alpha}) &
=m_{(0)}^{1}v_{(0)}^{\alpha}l_{\alpha_{(0)}},\\
J_{(0)}^{1}(\rho v^{i}l_{1}) &
=m_{(0)}^{1}v_{(0)}^{1}l_{1_{(0)}}+p_{(0)}/\gamma\cdot l_{1_{(0)}}.
\end{align}

The perturbative correction (first order) terms require some manipulation. The first order normal momentum flux is,
\begin{align}
J_{(1)}^{1}(\rho v^{i}l_{i}) &
=\left(m_{(0)}^{1}v_{(1)}^{i}+m_{(1)}^{1}v_{(0)}^{i}+p_{(1)}/\gamma\cdot
g_{(0)}^{i1}+p_{(0)}/\gamma\cdot
g_{(1)}^{i1}\right)l_{i_{(0)}}+\nonumber \\
& +\left(m_{(0)}^{1}v_{(0)}^{i}+p_{(0)}/\gamma\cdot
g_{(0)}^{i1}\right)l_{i_{(1)}}
\end{align}
with $g^{11}=1,\;\; g_{(0)}^{11}=1,\;\;\Rightarrow g_{(i)}^{11}=0\;\;\forall i\neq0$ and $\left(m_{(0)}^{1}v_{(0)}^{i}+p_{(0)}g_{(0)}^{i1}\right)=0$ yielding,
\begin{align}
J_{(1)}^{1}(\rho v^{i}l_{i}) &
=\left(m_{(0)}^{1}v_{(1)}^{i}+m_{(1)}^{1}v_{(0)}^{i}+p_{(1)}/\gamma\cdot
g_{(0)}^{i1}\right)l_{i_{(0)}}.
\end{align}
$J_{(1)}^{1}(\rho e)$ simplifies substantially due to $g^{i1}=0,\;\forall\; i\neq1$. Ultimately, the first order fluxes are
\begin{align}
J_{(1)}^{1}\begin{pmatrix}\rho\\
\rho v^{i}l_{i}\\
\rho e\\
\rho\lambda
\end{pmatrix}=\begin{pmatrix}m_{(1)}^{1}\\
\left(m_{(0)}^{1}v_{(1)}^{i}+m_{(1)}^{1}v_{(0)}^{i}+p_{(1)}/\gamma\cdot
g_{(0)}^{i1}\right)l_{i_{(0)}}\\
m_{(1)}^{1}e_{(0)}+m_{(0)}^{1}e_{(1)}+p_{(1)}/\gamma\cdot(m_{(0)}^{i}/\rho_{(0)})+p_{(0)}/\gamma\cdot\left(m^{i}/\rho\right)_{(1)}\\
m_{(1)}^{1}\lambda_{(0)}+m_{(0)}^{1}\lambda_{(1)}
\end{pmatrix}
\end{align}

The asymptotic expansion should be also performed in the source term $Q\approx Q_{(0)}+Q_{(1)}Z^{-1}$ . The calculation involves asymptotic expansion of the variables inside and outside the exponential function plus a Taylor expansion of the latter. Finally, the first and second order terms are
\begin{align}
Q_{(0)} &
=\rho_{(0)}k(1-\lambda_{(0)})e^{\left(-\theta_{a}/\left(p_{(0)}\Upsilon_{(0)}\right)\right)},\label{eq:source0}\\
Q_{(1)} &
=\left(\rho_{(1)}k(1-\lambda_{(0)})-\rho_{(0)}k\lambda_{(1)}\right)e^{\left(-\theta_{a}/\left(p_{(0)}\Upsilon_{(0)}\right)\right)}+\nonumber
\\
&
+\left[\rho_{(0)}k(1-\lambda_{(0)})\left(\Upsilon_{(1)}/\Upsilon_{(0)}+p_{(1)}/p_{(0)}\right)\theta_{a}/\left(p_{(0)}\Upsilon_{(0)}\right)\right]e^{\left(-\theta_{a}/\left(p_{(0)}\Upsilon_{(0)}\right)\right)}.
\end{align}
\subsection{Planar detonation}
\label{subsec:planar}
The Eq. \eqref{eq:J0} combined with \eqref{eq:flux0} and \eqref{eq:source0} constitute an equation system for \emph{planar} detonations. This system is equal to the known equations utilized to derive the classical results of the ZND theory of detonation, see \cite{Fic79}, \cite{Lee08}, \cite{Lee90}. The ZND theory provides an analytic expression for the pressure, velocity and specific volume profiles \cite{Lee90} as a function of the reaction progress variable $\lambda$,
\begin{align}
p & =a+(1-a)(1-b\beta\lambda)^{\frac{1}{2}},\; v=(1-p)(\gamma
M_{S})^{-1}+M_{s},\;\Upsilon=v/M_{s}.\label{eq:ZNDv}
\end{align}
The auxiliary variables appearing in Eq.\eqref{eq:ZNDv} are,
\begin{align}
& D=\tilde{D}/\tilde{c}_{s},\; M_{s}=\dfrac{(\gamma-1)D^{2}+2}{2\gamma
D^{2}-(\gamma-1)},\;\beta=\tilde{Q}\gamma/\tilde{c}_{s}^{2},\nonumber\\
& a=\dfrac{\gamma D^{2}+1}{2\gamma D^{2}-(\gamma-1)},\;
b=\dfrac{M_{s}^{2}2\gamma(\gamma-1)}{(1-a^{2})(\gamma+1)}.\;
\end{align}
Equations \eqref{eq:ZNDv} allow expressing the half reaction zone length as,
\begin{align}
\tilde{l_{c}} & =\tilde{c}_{s}k^{-1}{\textstyle
\int_{0}^{\frac{1}{2}}v(\lambda)(1-\lambda)^{-1}e^{\theta_{a}/(p(\lambda)\Upsilon(\lambda))}d\lambda.}
\end{align}
It is interesting to note that the whole system can be converted to the spatial formulation using the change of variables,
\begin{align}
X/Z=x & ={\textstyle \int_{0}^{\lambda}v(\zeta)/W(\zeta)d\zeta,}\\
dx/d\lambda & =v(\lambda)/W(\chi).\label{eq:change_var}
\end{align}
\subsection{Modeling}
\label{subsec:modeling}
A detonation can be considered as a small-thickness layer separating the fresh mixtures of the burned products. We propose the derivation of a three dimensional \emph{hydrodynamic} model in which the internal structure of the detonation as well as the chemical reaction is substituted by modified jump conditions. Conceptually, this construction asymptotically extends the \emph{simplest} planar stationary theory to three-dimensional non-stationary flow. 

In the derivation we consider two models for the detonation simultaneously, the \emph{hydrodynamic} and the \emph{reactive} model. Away from the consumption area, these models are identical. The jump conditions of the \emph{hydrodynamic} model, particularly the position and the amplitude of the discontinuity, are going to be determined from the internal structure of the detonation (\emph{reactive} model). Therefore, two sets of equations must be handled each corresponding to one of the models. Capital letters are used to designate the \emph{reactive} and lower-case for the \emph{hydrodynamic} model. The system of equations \eqref{eq:J0} and \eqref{eq:J1} may be the re-written as, 
\begin{align}
\partial_{X}\left(J_{(0)}^{1}\begin{pmatrix}\phi\\
\Phi
\end{pmatrix}\right) & =\begin{pmatrix}0\\
Q'_{(0)}(\Phi)
\end{pmatrix},\label{eq:mat_J0}\\
\partial_{X}\left(J_{(1)}^{1}\begin{pmatrix}\phi\\
\Phi
\end{pmatrix}\right) & =\begin{pmatrix}0\\
Q'_{(1)}(\Phi)-2HXQ'_{(0)}(\Phi)
\end{pmatrix}+2HJ_{(0)}^{1}\begin{pmatrix}\phi\\
\Phi
\end{pmatrix}-\nonumber \\
& -\left(\partial_{t}+\chi\right)\begin{pmatrix}\phi_{(0)}\\
\Phi_{(0)}
\end{pmatrix}-\left(g_{0}\right)^{-1/2}\partial_{x^{\alpha}}\left(\sqrt{g}_{(0)}J_{(0)}^{\alpha}\begin{pmatrix}\phi_{(0)}\\
\Phi_{(0)}
\end{pmatrix}\right).\label{eq:mat_J1}
\end{align}

In the detonation profile, see Figure \ref{fig:det_prof}, both models coincide in the initial, rarefaction and final states. The models exclusively differ in a thin zone surrounding the shock, i.e. between the \emph{hydrodynamic} discontinuity and the Chapman-Jouguet point, see Figure \ref{fig:det_detailed_prof}. In the area of appreciable chemical reaction the \emph{hydrodynamic} model is an extrapolation of the rarefaction wave. For simplicity, we set the origin of coordinates in the discontinuity of the \emph{hydrodynamic} model. Therefore, the Eq. \eqref{eq:mat_J0} and \eqref{eq:mat_J1} differ with respect to jump position, jump conditions and reaction source term. 

\vspace{0.5cm}

Figure 1.

\vspace{0.5cm}

Figure 2.

\vspace{0.5cm}

Subtracting \eqref{eq:mat_J0} from \eqref{eq:mat_J1} and rearranging we obtain,
\begin{align}
&\partial_{X}(J_{(0)}^{1}(\Phi)-J_{(0)}^{1}(\phi))=Q'_{(0)},\label{eq:both_J0}\\
&\partial_{X}(J_{(1)}^{1}(\Phi)-J_{(1)}^{1}(\phi))=Q'_{(1)}-2HXQ'_{(0)}(\phi)+\partial_{X}\left(2HX\left(J_{(0)}^{1}(\Phi)-J_{(0)}^{1}(\phi)\right)\right)-\nonumber
\\
&-\left(\partial_{t}+\chi\right)(\Phi_{(0)}-\phi_{(0)})-\left(g_{0}\right)^{-1/2}\partial_{x^{\alpha}}\left(\sqrt{g}_{(0)}\left(J_{(0)}^{\alpha}(\Phi)-J_{(0)}^{\alpha}(\phi)\right)\right).\label{eq:both_J1}
\end{align}

Equations \eqref{eq:both_J0} and \eqref{eq:both_J1} can be piecewise integrated from $-\infty$ to $\infty$. Taking into account the discontinuities and applying the \emph{Fundamental Calculus Theorem}, 
\begin{align}
&-\left[J_{(0)}^{1}(\Phi)\right]_{VN}+\left[J_{(0)}^{1}(\phi)\right]_{CJ}={\textstyle
\int_{-\infty}^{-\infty}Q_{(0)}(\Phi),\label{eq:both_int_J0}}\\
&-\left[J_{(1)}^{1}(\Phi)\right]_{VN}+\left[J_{(1)}^{1}(\phi)\right]_{CJ}={\textstyle
\int_{-\infty}^{-\infty}\left(Q_{(1)}(\Phi)-2HXQ_{(0)}(\Phi)\right)dX+}\nonumber\\
& +\partial_{X}\left({\textstyle
\int_{-\infty}^{-\infty}2HX\left(J_{(0)}^{1}(\Phi)-J_{(0)}^{1}(\phi)\right)dX}\right)-\left(\partial_{t}+\chi\right){\textstyle
\int_{-\infty}^{-\infty}\left(\Phi_{(0)}-\phi_{(0)}\right)dX-}\nonumber\\
& -\textstyle\int_{-\infty}^{-\infty}\left(g_{0}\right)^{-1/2}\partial_{x^{\alpha}}\left(\sqrt{g}_{(0)}\left(J_{(0)}^{\alpha}(\Phi)-J_{(0)}^{\alpha}(\phi)\right)\right)dX.\label{eq:both_int_J1}
\end{align}

We have designated with the index VN the \emph{reactive} discontinuity and with the index CJ the \emph{hydrodynamic}, by analogy of its conditions with the CJ point. The discontinuity in the \emph{reactive} model is a shock. Therefore, it is infinitely thin and the conditions applying for the shock discontinuity, see i.e. \cite{Lan87}, should also govern the curved case. Furthermore, the composition does not change across the discontinuity since the reaction starts at the high pressure side of the shock. Therefore,
\begin{align}
\left\{
\left[J^{1}(\Phi)\right]_{VN}=0,\;\left[J_{(0)}^{1}(\Phi)\right]_{VN}=0\right\}
\Rightarrow\left[J_{(1)}^{1}(\Phi)\right]_{VN}=0.
\end{align}
Substituting the condition $J_{(0)}^{1}(\Phi)=J_{(0)}^{1}(\phi)$ for $X\rightarrow-\infty$ into Eq. \eqref{eq:both_int_J0} we obtain for the vector $\theta=(\rho,\rho v^{i}l_{i},\rho e)$ 
\begin{align}
\left[J_{(0)}^{1}(\theta)\right]_{CJ} & =0,\;\;
J_{(0)}^{1}(\Theta)=J_{(0)}^{1}(\theta).
\end{align}
This has several implications. The mass flux satisfies $M_{(0)}^{j}-m_{(0)}^{j}=0$. In the moving curvilinear coordinates, $M_{(0)}^{\alpha}=m_{(0)}^{\alpha}=0$. Furthermore, the tangential component of momentum is continuous across the shock. Therefore, $J_{(0)}^{\alpha}(RV^{i}l_{i})=P_{(0)}/\gamma\cdot g_{(0)}^{i\alpha}l_{i_{(0)}}$ and thus, defining $\Pi=(P_{(0)}-p_{(0)})/\gamma$ we obtain $J_{(0)}^{\alpha}(RV^{i}l_{i})-J_{(0)}^{\alpha}(\rho v^{i}l_{i})=\Pi g_{(0)}^{i\alpha}l_{i_{(0)}}$. Finally, the normal component of the momentum flux, $J_{(0)}^{1}(RV^{i}l_{i})l^{1}$, is identical in the \emph{reactive} and CJ \emph{hydrodynamic} model. Therefore, $\Pi=M_{(0)}^{1}V_{(0)}^{1}-m_{(0)}^{1}v_{(0)}^{1}$. 

The Eq. \eqref{eq:both_int_J1} can be simplified applying the previous considerations. For continuity, momentum and energy the first two terms in the RHS of \eqref{eq:both_int_J1} vanish. Moreover, the last term of the RHS cancels for continuity due to vanishing transverse momentum. For the continuity Eq. \eqref{eq:both_int_J1} reduces to 
\begin{align}
\left[m_{(1)}^{1}\right]_{CJ} &
=-\left(\partial_{t}+\chi\right){\textstyle
\int_{-\infty}^{\infty}(R_{(0)}^{1}-\rho_{(0)}^{1})dX\label{eq:both_cont_simp}.}
\end{align}
The simplifications in the momentum equation are,
\begin{align}
&\left[\left(m_{(0)}^{1}v_{(1)}^{i}+m_{(1)}^{1}v_{(0)}^{i}+g_{(0)}^{i1}p_{(1)}\gamma^{-1}\right)l_{i_{(0)}}\right]_{CJ}=-{\textstyle
\int_{-\infty}^{\infty}g_{0}^{-\frac{1}{2}}\partial_{x^{\alpha}}(g_{(0)}^{\frac{1}{2}}\Pi
g_{(0)}^{i\alpha}l_{i_{(0)}})dX-}\nonumber\\
& -{\textstyle
\int_{-\infty}^{\infty}\left(\partial_{t}+\chi\right)(R_{(0)}V_{(0)}^{i}-\rho_{(0)}v_{(0)}^{i})l_{i_{(0)}}dX\label{eq:both_imp_simp},}
\end{align}
those of the energy equation are,
\begin{align}
& \left[m_{(1)}^{1}e_{(0)}+m_{(0)}^{1}e_{(1)}+p_{(1)}/\gamma\cdot
g_{(0)}^{i1}\cdot m_{(0)}^{i}/\rho_{(0)}+p_{(0)}/\gamma\cdot
g_{(0)}^{i1}\left(m^{i}/\rho\right)_{(1)}\right]_{CJ}=\nonumber \\
& =-{\textstyle
\int_{-\infty}^{\infty}\left(\partial_{t}+\chi\right)\left(R_{(0)}E_{(0)}-\rho_{(0)}e_{(0)}\right)dX\label{eq:both_ener_simp},}
\end{align}
and finally those of the species equation yield,
\begin{align}
&\left[m_{(1)}^{1}\lambda_{(0)}+m_{(0)}^{1}\lambda_{(1)}\right]_{CJ}={\textstyle
\int_{-\infty}^{+\infty}\left(Q_{(1)}(RL)-2HXQ_{(0)}(RL)\right)dX}-\nonumber\\
& -{\textstyle
\int_{-\infty}^{+\infty}\left(\partial_{t}+\chi\right)\left(R_{(0)}L_{(0)}-\rho_{(0)}\lambda_{(0)}\right)dX.}
\end{align}
We may define $I_{R}=\int_{-\infty}^{\infty}\left(R_{(0)}-\rho_{(0)}\right)dX$, $I_{\sigma}=\int_{-\infty}^{\infty}\Pi dX$ and $I_{\Sigma}=\int_{-\infty}^{\infty}(R_{(0)}E_{(0)}-\rho_{(0)}e_{(0)})dX$. The equations, $R_{(0)}V_{(0)}^{i}-\rho_{(0)}v_{(0)}^{i}=u_{(0)}^{i}(R_{(0)}-\rho_{(0)})$ and $(g_{0})^{-1/2}\cdot\partial_{x^{\alpha}}(\sqrt{g}_{(0)}g_{(0)}^{i\alpha}l_{i_{(0)}})=-2Hl_{1}+(g_{0})^{-1/2}\partial_{x^{\alpha}}(\sqrt{g}_{(0)}g_{(0)}^{\beta\alpha}l_{\beta_{(0)}})$, see \cite{Cla03}, allow manipulating \eqref{eq:both_cont_simp}, \eqref{eq:both_imp_simp} and \eqref{eq:both_ener_simp} to obtain 
\begin{align}
&\left[m_{(1)}^{1}\right]_{CJ}=-\left(\partial_{t}+\chi\right)I_{R},\label{eq:last_both_den}\\
&\left[\left(m_{(0)}^{1}v_{(1)}^{i}+m_{(1)}^{1}v_{(0)}^{i}+p_{(1)}/\gamma\cdot
g_{(0)}^{i1}\right)l_{i_{(0)}}\right]_{CJ}=\nonumber \\
&=-\left(\partial_{t}+\chi\right)u_{(0)}^{i}l_{i(0)}I_{R}-\left(2Hl_{1}+g^{\alpha\beta}l_{\beta}\partial_{x^{\alpha}}\right)I_{\sigma},\\
& \left[m_{(1)}^{1}e_{(0)}+m_{(0)}^{1}e_{(1)}+p_{(1)}/\gamma\cdot
g_{(0)}^{i1}\cdot m_{(0)}^{i}/\rho_{(0)}+p_{(0)}/\gamma\cdot
g_{(0)}^{i1}\left(m^{i}/\rho\right)_{(1)}\right]_{CJ}=\nonumber \\
& =-\left(\partial_{t}+\chi\right)I_{\Sigma}.
\end{align}

We combine the leading order jump conditions with the perturbative corrections and decompose the jump condition for the momentum in normal and tangential components. The result obtained is 
\begin{equation}
\begin{bmatrix}
m^{1}\\
m^{1}v^{1}+p/\gamma\\
m^{1}v^{\beta}\\
m^{1}e+m^{1}p/\rho\gamma
\end{bmatrix}_{CJ}=-\left((\partial_{t}+\chi)
\begin{pmatrix}I_{R}\\
u_{(0)}^{1}I_{R}\\
u_{(0)}^{\beta}I_{R}\\
I_{\Sigma}
\end{pmatrix}+\begin{pmatrix}0\\
2HI_{\sigma}\\
g^{\alpha\beta}\partial_{x^{\alpha}}I_{\sigma}\\
0
\end{pmatrix}\right)Z^{-1}+o(Z^{-1}).\label{eq:total_all_mat}
\end{equation}

The first row of the Eq. \eqref{eq:total_all_mat} shows that the normal mass flux $m^{1}$ experiences $O(Z^{-1})$ variations in the detonation structure. This represents the excess of mass in the \emph{hydrodynamic} model compared to the \emph{reactive} model. Following the methodology described in \cite{Cla03}, we define without loss of generality the position of the artificial discontinuity in the \emph{hydrodynamic} model requiring identical normal mass flux in the fresh and burned mixtures. A continuous mass flux across the discontinuity surface is obtained. The condition $\left[m^{1}\right]_{CJ}=0$ requires $I_{R}=0$ where, 
\begin{align}
I_{R}= & {\textstyle
\int_{-\infty}^{0}(R_{(0)}-\rho_{(0)})dX+{\textstyle
\int_{0}^{x_{VN}}(R_{(0)}-\rho_{(0)})dX+}}\nonumber\\
& +{\textstyle \int_{x_{VN}}^{x_{CJ}}(R_{(0)}-\rho_{(0)})dX+{\textstyle
\int_{x_{CJ}}^{\infty}(R_{(0)}-\rho_{(0)})dX\label{eq:IR_total}.}}
\end{align}
The areas under the curves in Figure \ref{fig:det_prof} and \ref{fig:det_detailed_prof} delimited by the \emph{hydrodynamic} discontinuity and the Chapman-Jouguet point cancel. Therefore, the first and last term of the RHS of Eq. \eqref{eq:IR_total} cancel. The difference of slopes, see Figure \ref{fig:det_detailed_prof}, between rarefaction and consumption curves is asymptotically large. In stretched coordinates, the slope of the hydrodynamic model is zero at leading order and does not affect the integral $I_{R}$ in the consumption area. After some manipulation we obtain, 
\begin{align}
I_{R} & \approx{\textstyle
\int_{x_{VN}}^{x_{CJ}}(R_{(0)}-\rho_{CJ})dX-(\rho_{CJ}-\rho_{0})X_{VN},}
\end{align}
as can be graphically confirmed in the Figures \ref{fig:det_prof} and \ref{fig:det_detailed_prof}. The position of the detonation shock relative to the artificial discontinuity is 
\begin{align}
X_{VN} & \approx\left(\rho_{CJ}-\rho_{0}\right)^{-1}{\textstyle
\int_{X_{VN}}^{X_{CJ}}\left(R_{(0)}-\rho_{CJ}\right)dX\label{eq:possition}.}
\end{align}
By similar considerations we find,
\begin{align}
I_{\sigma} & =\gamma^{-1}{\textstyle
\int_{X_{VN}}^{X_{CJ}}\left(P_{(0)}-p_{(0)}\right)dX\approx\gamma^{-1}\left(I_{1}-\left(p_{CJ}-p_{0}\right)X_{VN}\right),}
\end{align}
where $I_{1}=\int_{X_{VN}}^{X_{CJ}}\left(P_{(0)}-p_{CJ}\right)dX$. The terms inside integral $I_{1}$ may be reformulated changing the variables according to Eq.\eqref{eq:change_var}. Now, the more convenient composition formulation is recovered, and thus $I_{1}=Z \textstyle \int_{0}^{1}\left(P(\lambda)-p_{CJ}\right) u(\lambda) \cdot \left(r(\lambda)\right)^{-1} d\lambda = ZI_{1}'$. Similarly Eq. \eqref{eq:possition} can be expressed in the composition formulation yielding, 
\begin{align}
x_{VN}\approx & \dfrac{{\textstyle
\int_{0}^{1}(R_{(0)}^{1}(\lambda)-\rho_{CJ})u(\lambda)\left(r(\lambda)\right)^{-1}d\lambda}}{\rho_{CJ}-\rho_{0}}=\dfrac{I_{\rho}}{\rho_{CJ}-\rho_{0}}.
\end{align}
Finally $I_{\sigma}$ is reformulated, 
\begin{align}
I_{\sigma}= &
\gamma^{-1}(I_{1}'-(p_{CJ}-p_{0})(\rho_{CJ}-\rho_{0})^{-1}I_{\rho})Z,
\end{align}
and $I_{\Sigma}$ is rescaled $I_{\Sigma}=I'_{\Sigma}Z$. The Eq. \eqref{eq:total_all_mat} is rewritten as 
\begin{equation}
\begin{bmatrix}
m^{1}\\
m^{1}v^{1}+p/\gamma\\
m^{1}v^{\beta}\\
m^{1}e+m^{1}p/\rho\gamma
\end{bmatrix}_{CJ}=-(\partial_{t}+\chi)
\begin{pmatrix}
0\\
0\\
0\\
I'_{\Sigma}
\end{pmatrix}
-
\begin{pmatrix}
0\\
2H\gamma^{-1}\\
g^{\alpha\beta}\gamma^{-1}\partial_{x^{\alpha}}\\
0
\end{pmatrix}
\left(I_{1}'-\dfrac{p_{CJ}-p_{0}}{\rho_{CJ}-\rho_{0}}I_{\rho}\right)+o(Z^{-1}).\label{eq:total_all_mat2}
\end{equation}
The integrals $I_{1}$ and $I_{\rho}$ can easily be understood considering the Figure \ref{fig:det_detailed_prof}. Both integrals represent areas contained between the detonation curve and the horizontal CJ conditions. $I_{1}$ and $I_{\rho}$ depend on the chemistry model. Even numerical evaluation for complex chemistry is possible. In the present work, the explicit expressions, Eq. \eqref{eq:ZNDv}, were selected in the analysis. Obviously, chemical conversion will not fully complete in a layer of finite thickness. This does not contradict our assumption of a consumption layer of small finite thickness, since the chemical conversion becomes exponentially small for $\lambda\rightarrow1$. Accordingly the integrals $I_{1}$ and $I_{\rho}$ converge for $\lambda\rightarrow1$. Note that, analogous to references \cite{Cla03} and \cite{Cla03b} the stretch $\chi$ can be calculated as 
\begin{align}
\chi &
=\left.|\nabla\rho|^{-1}\nabla\cdot\left(|\nabla\rho|\vec{u}_{(0)}\right)\right|_{X=0^{+}}.\label{eq:stretch}
\end{align}
Here the stretch experiences a jump at the \emph{hydrodynamic} discontinuity surface. Thus, the evaluation is taken on the high pressure side of the interface, a fact that is emphasized by the symbol $+$ in Eq. \eqref{eq:stretch}.

\subsection{Final jump conditions}
\label{subsec:fjump} 

A clearer understanding of the jump conditions may be obtained re-writing Eq. \eqref{eq:total_all_mat2} in dimensional form. $I'_{1}$ and $I_{\rho}$ were transformed including the reference values $\tilde{l_{f}}$, $\tilde{p}_{s}=\tilde{\rho}_{s}\tilde{c}_{s}^{2}/\gamma$ and $\tilde{\rho}_{s}$ in the kernel and dummy variable of the integrals. Eq. \eqref{eq:total_all_mat_dim} reads in dimensional notation 
\begin{equation}
\begin{bmatrix}
\tilde{m}^{1}\\
\tilde{m}^{1}\tilde{v}^{1}+\tilde{p}\\
\tilde{m}^{1}\tilde{v}^{\beta}\\
\tilde{m}^{1}\tilde{e}+\tilde{v}^{1}\tilde{p}
\end{bmatrix}_{CJ}\approx-(\partial_{\tilde{t}}+\tilde{\chi})\begin{pmatrix}0\\
0\\
0\\
\tilde{I}'_{\Sigma}
\end{pmatrix}
-
\begin{pmatrix}
0\\
2\tilde{H}\\
g^{\alpha\beta}\partial_{\tilde{x}^{\alpha}}\\
0
\end{pmatrix}
\left(\tilde{I}_{1}'-\dfrac{\tilde{p}_{CJ}-\tilde{p}_{0}}{\tilde{\rho}_{CJ}-\tilde{\rho}_{0}}\tilde{I}_{\rho}\right).\label{eq:total_all_mat_dim}
\end{equation}

It is interesting to note, that the system of equations representing the \emph{planar} detonation \eqref{eq:ZNDv} is, for a gas of known composition, a mono-parametric system dependent on $D$. The system represented by \eqref{eq:total_all_mat_dim} depends not only on $D$, but also on curvature $\tilde{H}$ and on stretch $\tilde{\chi}$. 

The velocity of the detonation $D$ also suffers a change due to curvature that can be expressed through an asymptotic expansion of the form $D=D_{(0)}+D_{(1)}Z^{-1}$. The velocity of the detonation must be derived from the species equation. This derivation exceeds the scope of this paper and will be included in a forthcoming publication. 

\subsection{Virtual surface tension}
\label{subsec:surf_ten} 

The normal momentum Eq. \eqref{eq:total_all_mat2} is discontinuous across the jump with linear proportionality on curvature. Following the arguments in \cite{Cla03}, an analogy can be established with an interface separating two immiscible fluids \cite{Lan87} in order to calculate the virtual surface tension of the detonation. At the interface,
\begin{align}
\left[\tilde{p}-\sigma_{nn}\right]= & 2\tilde{H}\alpha,
\end{align}
where $\alpha$ represents the surface tension. Our initial scale considerations suggest that the tangential stress is negligible. Allowing for mass transfer across the surface (e.g. due to evaporation) the formula must be modified 
\begin{align}
\left[\tilde{m}^{1}\tilde{v}^{1}+\tilde{p}\right]= & 2\tilde{H}\alpha.
\end{align}
We may, identify terms with eq. \eqref{eq:total_all_mat_dim} to obtain 
\begin{align}
\alpha= &
-\left(\tilde{I}_{1}'-\dfrac{\tilde{p}_{CJ}-\tilde{p}_{0}}{\tilde{\rho}_{CJ}-\tilde{\rho}_{0}}\tilde{I}_{\rho}\right)=-\tilde{l_{f}}\left(I'_{1}\dfrac{\rho_{s}c_{s}^{2}}{\gamma}-\dfrac{\tilde{p}_{CJ}-\tilde{p}_{0}}{\tilde{\rho}_{CJ}-\tilde{\rho}_{0}}\rho_{s}I_{\rho}\right)=\tilde{l_{f}}\alpha_{0}.
\end{align}
The coefficient of surface tension in an infinitely thin gas-dynamic discontinuity equivalent to a detonation is equal to the difference between the integral of the pressure between the CJ and the VN points minus the integral of the density between the same integration limits normalized by a factor.

 The evaluation of $\alpha_{0}$ and the partial factors $\alpha_{01}=-I'_{1}\rho_{s}c_{s}^{2}/\gamma$ and $\alpha_{02}=(\tilde{p}_{CJ}-\tilde{p}_{0})(\tilde{\rho}_{CJ}-\tilde{\rho}_{0})^{-1}\rho_{s}I_{\rho}$ is included in the Figure \ref{fig:over_fact_1} for a gas of the indicated characteristics and different initial pressures $p_{0}$. The surface tension exhibits an inverse proportionality to the initial pressure. The existence of a minimum is due to the sum of the $\alpha_{01}$ and $\alpha_{02}$ factors (dashed lines) that combined creates the final dependency.

\vspace{0.5cm}

Figure 3.

\vspace{0.5cm}

Figure \ref{fig:over_fact_2} shows the dependence of the surface stress on the fuel concentration. An increase on the fuel concentration strongly increases the surface tension of the equivalent jump. 

\vspace{0.5cm}

Figure 4.

\vspace{0.5cm}

The existence of virtual surface tension has strong implications for the stability of the detonation. In this sense, Eq. \eqref{eq:total_all_mat_dim} shows that the tangential momentum must not be continuous across the interface. The derivative of the surface tension, RHS of Eq. \eqref{eq:total_all_mat_dim}, has an analogous meaning to the Marangoni forces. A detonation may also propagate isotropically in a medium of constant composition forming cellular structures \cite{Lee08}, \cite{Fic79}. The RHS of the tangential impulse in Eq.\eqref{eq:total_all_mat_dim} will cancel. The normal momentum component in Eq. \eqref{eq:total_all_mat_dim} can be exploited to show that the mean curvature is constant at leading order for a complete cellular structure. However, the curvature will change in time through the derivative dependence of the energy jump condition.

\section{Conclusions}
\label{sec:conclu} 

In this paper we have derived a model which describes the detonation phenomenon as a gas-dynamic discontinuity. Our model was derived from the Euler equations. Modified Rankine-Hugoniot conditions link the fluid fields on the two sides of the detonation surface. The amplitudes of the jump depend of the precise location of the discontinuity related to the detonation structure. The position is defined by a density integral which yields vanishing of the excess surface mass, leading to the continuity of the mass flux across the flame. The continuity of the mass flux simplifies the conditions substantially and allows for physical interpretation. An additional pressure jump proportional to the curvature appeared. This may be interpreted as virtual surface tension. The jump conditions obtained are valid for arbitrary chemistry. We plan to implement numerical codes for the computation of detonations based on our model.

\section*{Acknowledgment}
This research was partially supported by DFG grant SFB606.
\bibliographystyle{model1-num-names}

\bibliography{bibl.bib}

\newpage
\begin{center}
  \begin{figure}[h]
    \centering 
    \includegraphics*
    {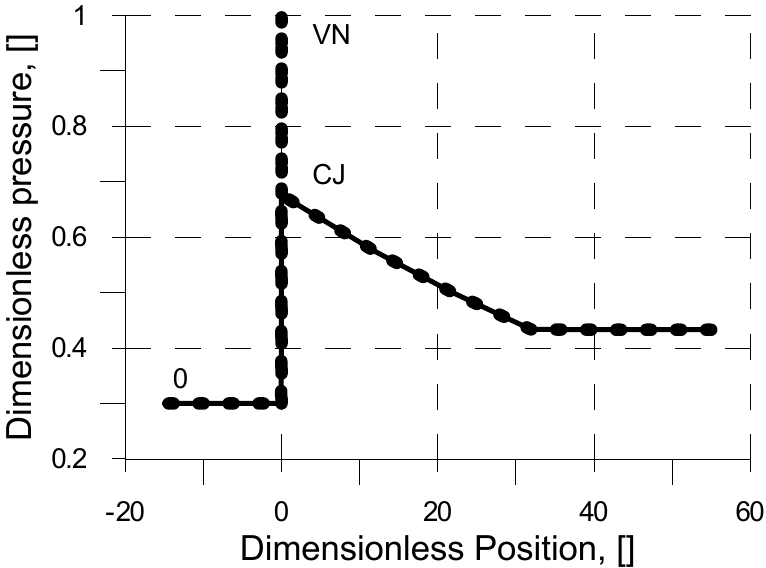} 
    \caption{Profiles of \emph{hydrodynamic} (solid) and \emph{reactive} (dashed) detonation models. Global view. Profile obtained with ZND theory coupled with rarefaction wave, for a gas of characteristics $p_{0}=100\, kPa$, $\rho_{0}=1\, kg/m^{3}$, $Q=0.1\, MJ$, $\gamma=1.4$, $k=1\,10^{5}s^{-1}$, $E/R_{g}=10000\, K$. \emph{0} designates normal status, \emph{VN} von Neumann peak, \emph{CJ} Chapman-Jouget point.}
    \label{fig:det_prof}
  \end{figure}
\end{center}
\newpage
\begin{center}
  \begin{figure}[h]
    \centering 
    \includegraphics
    {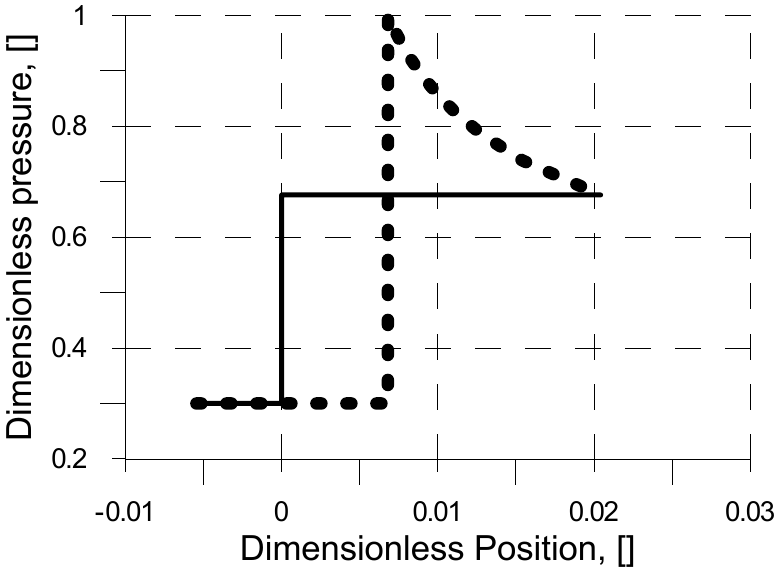} 
    \caption{Profiles of \emph{hydrodynamic} (solid) and \emph{reactive} (dashed) detonation models. Detailed area. Profiles obtained with the same conditions as in Figure \ref{fig:det_prof}}
    \label{fig:det_detailed_prof}
  \end{figure}
\end{center}

\newpage
\begin{center}
  \begin{figure}[h]
    \centering 
    \includegraphics*
    {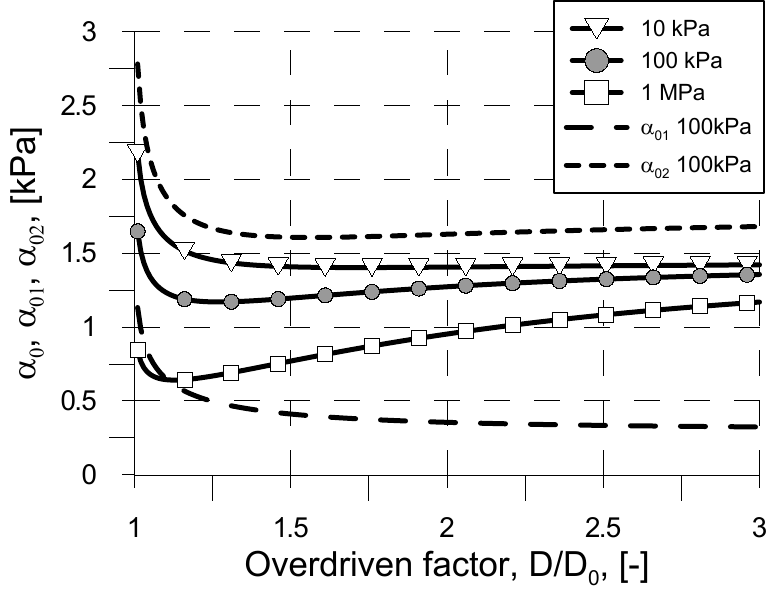} 
    \caption{Dependence of $\alpha_{0}$ factor to the degree of over-driven detonation  for a gas of characteristics $\rho_{0}=1\, kg/m^{3}$, $Q=0.1\, MJ$, $\gamma=1.4$,$k=1\,10^{5}s^{-1}$, $E/R_{g}=10000\, K$ obtained for different pressures}
    \label{fig:over_fact_1}
  \end{figure}
\end{center}

\newpage

\begin{center}
  \begin{figure}[h]
    \centering 
    \includegraphics*
    {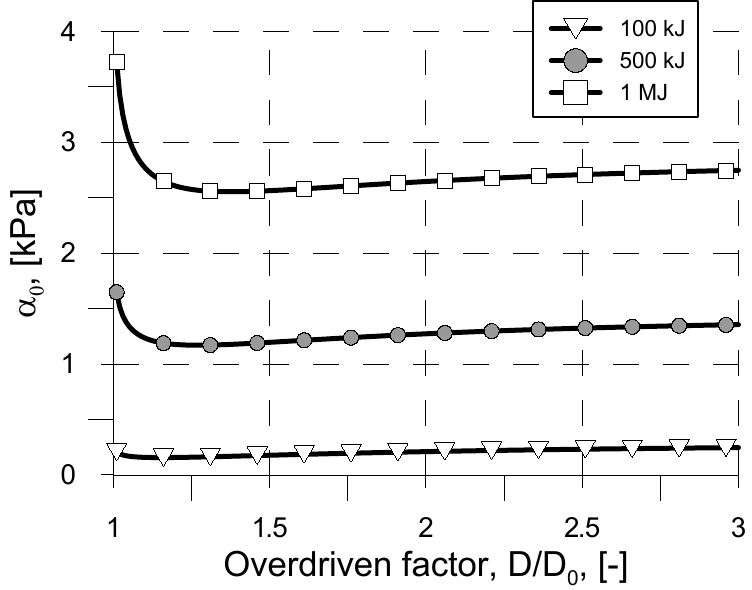} 
    \caption{Dependence of $\alpha_{0}$ factor to the degree of over-driven detonation for a gas of characteristics $p_{0}=100\, kPa$, $\rho_{0}=1\, kg/m^{3}$, $\gamma=1.4$,$k=1\,10^{5}s^{-1}$, $E/R_{g}=10000\, K$ obtained for different enthalpies of formation} 
    \label{fig:over_fact_2}
  \end{figure}
\end{center}

\newpage

List of captions:

\vspace{1.0 cm}

Figure 1: 
Profiles of \emph{hydrodynamic} (solid) and \emph{reactive} (dashed) detonation models. Global view. Profile obtained with ZND theory coupled with rarefaction wave, for a gas of characteristics $p_0=100\, kPa$,  $\rho_0=1\, kg/m^3$,  $Q=0.1\,MJ$,  $\gamma=1.4$,  $k=1\,10^5 s^{-1}$,  $E/R_g=10000\, K$. \emph{0} designates normal status, \emph{VN} von Neumann peak, \emph{CJ} Chapman-Jouget point. 
\vspace{1.0 cm}

Figure 2: 
Profiles of \emph{hydrodynamic} (solid) and \emph{reactive} (dashed) detonation models. Detailed area. Profiles obtained with the same conditions as in Figure \ref{fig:det_prof}. 

\vspace{1.0 cm}

Figure 3: 
Dependence of $\alpha_0$ factor to the degree of overdriven detonation for a gas of characteristics  $\rho_0=1\,  kg/m^3$,  $Q=0.1\,MJ$,  $\gamma=1.4$, $k=1\,10^5 s^{-1}$, $E/R_g=10000\, K$ obtained for different pressures. 

\vspace{1.0 cm}

Figure 4: 
Dependence of $\alpha_0$ factor to the degree of overdriven detonation for a gas of characteristics $p_0=100\, kPa$, $\rho_0=1\, kg/m^3$, $\gamma=1.4$, $k=1\,10^5 s^{-1}$, $E/R_g=10000\, K$ obtained for different enthalpies of formation.

\end{document}